\documentclass[ floatfix,twocolumn]{revtex4}
\usepackage{amsmath}
\usepackage{color}
\usepackage{amssymb}
\usepackage{graphicx}
\usepackage[utf8]{inputenc}
\usepackage[english, russian]{babel}
\usepackage{textcomp}
\usepackage{color}
\usepackage{esint}

\begin{document}

\title{ 
Superconducting states in metals with toroidal ordering}

\author{V.P.Mineev}
\affiliation{Landau Institute for Theoretical Physics, 142432 Chernogolovka, Russia}

\begin{abstract}

The paper presents a theory of superconducting states in metals with toroidal symmetry.

\end{abstract}
\date{\today}
\maketitle

\section{Introduction}

The possibility of a magnetoelectric effect
in some antiferromagnetic crystals was first noted in the textbook by L.D. Landau and E.M. Lifshitz "Electrodynamics of Continuous Media" in 1958  \cite{LL1958}. This phenomenon is associated with a special symmetry of the material, in which the time reversal symmetry
$R$ and the inversion symmetry
$I$ are violated,
although the product $RI$ is still a symmetry operation.
The thermodynamic potential of such a solid must contain terms proportional to the product of the first powers of the components of the electric and magnetic fields $(\Phi\sim EH )$. I.E. Dzyaloshinskii \cite{Dzyal1959} gave a specific example of this type of antiferromagnet, namely, Cr$_2$O$_3$, where, unlike ordinary two - or more sublattice antiferromagnets, the magnetic moments of Cr are in the same crystallographic unit cell. The point symmetry group of Cr$_2$O$_3$
${\bf D}_{3d}({\bf D}_3)=(E, C_3,C_3^2,3u_2,3\sigma_dR,2S_6R,IR)$ contains the product of time and space inversion, but does not include these operations separately. Soon after, D.N. Astrov \cite {Astrov1960} discovered the magnetoelectric effect in this material.

Some of the magnetoelectric antiferromagnets exhibit optical nonreciprocity, which is the difference in reflectivity of left and right circularly polarized light (the Kerr effect)
despite the absence of a net magnetic moment \cite{Brown1963,Hornreich1968}. In Cr$_2$O$_3$ the effect was observed experimentally by Krichevtsov et al. \cite{Krichevtsov1993,Krichevtsov1996}. 
New interest in these type of phenomena arose after spin-flip neutron scattering experiments on the underdoped high-temperature superconducting compound YBa$_2$Cu$_3$O$_{ 6+x}$ \cite{Fauque2006}, indicating a violation of time reversal that occurs without appearance 
spontaneous magnetisation and without  loss of translational symmetry.
This was followed by the discovery of the Kerr effect on the same material by A. Kapitulnik's group \cite{Xia2008}. The experiments were interpreted by C.Varma in terms of the "loop current model" (see recent review \cite{Varma2020}).

At the same time, these experimental observations raised the problem of describing the superconducting state in materials with broken time reversal and inversion symmetry, but possessing symmetry with respect to the product of these operations, which are now usually called substances with toroidal order \cite{Hayami2014}.
Several issues related to this problem have been considered in
\cite{Kivelson2008,Allais2012,Agterberg2023} in the framework of a single-band 2D model with d-wave superconducting ordering, and in the framework of a 1D model developed by S. Sumita and Y. Yanase
\cite{Yanase2016}. Here we present a more general treatment of superconductivity in antiferromagnetic materials with a Cr$_2$O$_3$-like structure, with magnetic moments located within a single unit cell, so that time and space inversion are broken, but the total magnetic moment is zero and symmetry with respect to the product of time inversion and time reversal operations holds. Several examples of structures of this type have been discussed by J. Orenstein \cite{Orenstein2011} in connection with the possible ordering  of magnetic moments created by loop currents in YBCO.

\begin{figure}
\includegraphics
[height=.2\textheight]
{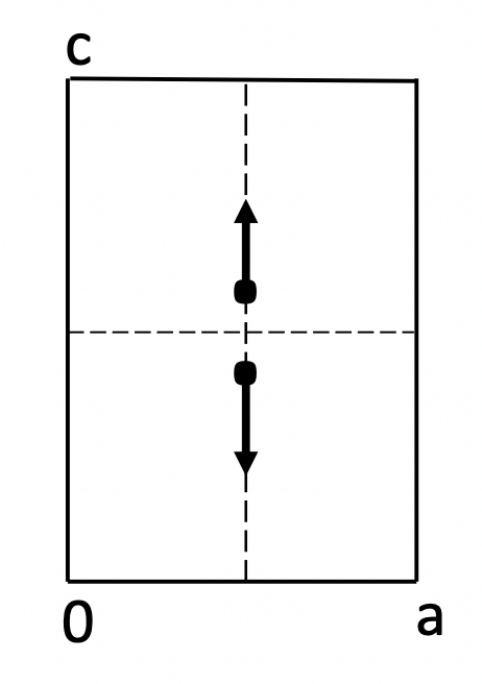}
 \caption{
The $y=b/2$ plane  of elementary cell of an antiferromagnet with point  symmetry  group ${\bf D}_{2h}({\bf D}_2)$. }
\end{figure}

Specific examples of superconductors of this type are currently unknown, although there are metals and semiconductors with toroidal magnetic ordering. See, for example, the paper by H.Amitsuka et al \cite{Amitsuka2018} about UNi$_4$B and O.Fedchenko et al
\cite{Fedchenko2022} on Mn$_2$Au. With hope for future experimental discoveries, we develop here a symmetry approach to the description of superconductivity in materials of this type. This will be done on example of  orthorhombic antiferromagnet (see Fig. 1) with the point symmetry group ${\bf D}_{2h}({\bf D}_2)$, which allows  demonstrate the specific features of the theory of superconducting states in substances with toroidal order.
Examples of other toroidal antiferromagnets with other point symmetry groups are listed in the Appendix.

The paper is organised as follows. The general symmetry properties are described in the next chapter. The third chapter is devoted to the development of the BCS-type theory in toroids. The results are collected in the Conclusion, where prospects for future research are also indicated. The Appendix demonstrates specific features of superconducting toroids with a point symmetry group different from that described in the main text.

\section{Structure of superconducting states}

We will consider 3D metal with orthorhombic crystal structure with point symmetry group ${\bf D}_{2h}({\bf D}_2) $ including the following operations 
\begin{equation}
E,C_{2x},C_{2y},C_{2z},R\sigma_x,R\sigma_y,R\sigma_z,RI.
\end{equation}
Each   electron energy band consists of sum of   even and odd  functions of components of the wave vector ${\bf k}$  
\begin{eqnarray}
\varepsilon_{\bf k}=\varepsilon^e_{\bf k}+\varepsilon^o_{\bf k},~~~~~~\varepsilon^e_{\bf k}=f(k_x^2,k_y^2,k_z^2),\\\varepsilon^o_{\bf k}=\gamma\sin k_xa\sin k_yb\sin k_zc~~~~~~~
\label{epsilon}
\end{eqnarray}
invariant in respect of all operations of the point symmetry  group ${\bf D}_{2h}({\bf D}_2) $.
Corresponding Fermi surface determined by equation
\begin{equation}
\varepsilon_{\bf k}=\varepsilon_F,
\end{equation}
is assymetrical inasmuch as   $\varepsilon_{\bf k} \ne \varepsilon_{-{\bf k}}$.

Hamiltonian in the Schr\"odinger equation for an electron in such a metal commutes with the product of time and space inversion operations $RI$.
 This means that to each energy $\varepsilon_{\bf k}$ corresponds two spinor eigen-functions $\psi_{\bf k\alpha}({\bf r})$
 and $RI\hat \psi_{\bf k\alpha}({\bf r})$ orthogonal  each other.

 The magneto-electric part of thermodynamic potential has the following form $$\Phi_{ME}=-\alpha_xE_xH_x-\alpha_yE_yH_y-\alpha_zE_zH_z.$$
 
 The orthorhombic group $D_{2h}(D_2)$ has four one-dimensional representations. The absence symmetry in respect inversion means that all superconducting 
 states
 consists of sum singlet and triplet states. The corresponding order parameters  of superconducting states  have the following form
 \begin{equation}
 \Delta_{\alpha\beta}({\bf k})=\Delta\Phi_{\alpha\beta}({\bf k})=\Delta \left [ \phi^s_{\bf k}i\sigma^y_{\alpha\beta}+(\mbox{\boldmath$\phi$}^t_{\bf k}\mbox{\boldmath$\sigma$}_{\alpha\gamma})i\sigma^y_{\gamma\beta}\right ].
\label{op}
 \end{equation}
Here, $ \mbox{\boldmath$\hat\sigma$}=(\hat\sigma^x,\hat\sigma^y,\hat\sigma^z)$ are the Pauli matrices. The functions $\phi^s_{\bf k}$ and 
$\mbox{\boldmath$\phi$}^t_{\bf k}$ for all four representations $\Gamma$=A,B$_3$,B$_2$,B$_1$ are presented in the table

\bigskip

 \begin{tabular}{|c|c|c|}
\hline
 $\Gamma$& $\phi^s_{\bf k}$&$\mbox{\boldmath$\phi$}^t_{\bf k}$ \\
 \hline
 ~A~& ~$a\hat k_x^2+ b\hat k_y^2+c\hat k_z^2$~&~$ i( \hat k_x\hat x+\hat k_y\hat y+\hat k_y\hat z)$~\\
 \hline
 ~B$_3$~&~$i \hat k_y\hat k_z$~&~$ \hat k_z\hat y+\hat k_y\hat z$~\\
 \hline
  ~B$_2$~&~$i \hat k_x\hat k_z$~&~$ \hat k_z\hat x+\hat k_x\hat z$~\\
  \hline
 ~B$_1$~&~$i \hat k_x\hat k_y$~&~$ \hat k_y\hat x+\hat k_x\hat y$~\\
  \hline
 \end{tabular}
 
\bigskip 

Here, it is necessary to keep in mind that each term in each cell may include some amplitude (a real numerical factor or a real function of  squares of  unit wave vector components $\hat{\bf k}={\bf k}/|{\bf k}|$). For example, the vector function for the triplet state $A$ is generally equal to
$i( f\hat k_x\hat x+g\hat k_y\hat y+h\hat k_y\hat z)$ where $f=f(\hat k_x^2,\hat k_y^2,\hat k_z^2)$ etc.

\section{BCS  theory for ferroelectric metals} 

The Hamiltonian including interaction of electrons with opposite momenta is
\begin{eqnarray}
H=H_0+H_{int}=\sum_{\bf k}(\xi_{\bf k}+\varepsilon^o_{\bf k})a^+_{{\bf k}\alpha}a_{{\bf k}\alpha}\nonumber\\
+\frac{1}{2}\sum_{{\bf k},{\bf k}^\prime}V_{\alpha\beta,\lambda\mu}({\bf k},{\bf k}^\prime)
a^+_{-{\bf k}\alpha}a^+_{{\bf k}\alpha}a_{{\bf k}^\prime\lambda}a_{-{\bf k}^\prime\mu}.
\label{H}
\end{eqnarray}
Here 
\begin{equation}
\xi_{\bf k}=\varepsilon^e_{\bf k}-\mu
\end{equation}
and  in the pairing interaction 
\begin{equation}
V_{\alpha\beta,\lambda\mu}({\bf k},{\bf k}^\prime)=-V_\Gamma\Phi_{\alpha\beta}({\bf k})\Phi^\dagger_{\lambda\mu}({\bf k}^\prime)
\end{equation}
was left only term related to irreducible representation $\Gamma$ corresponding  to superconducting state with maximal critical temperature. 
The normal part of Hamiltonian can be rewritten as
\begin{eqnarray}
H_0=\frac{1}{2}\sum_{\bf k}(\xi_{\bf k}+\varepsilon^o_{\bf k})a^+_{{\bf k}\alpha}a_{{\bf k}\alpha}~~~~~~~~\nonumber\\
-\frac{1}{2}\sum_{\bf k}(\xi_{-\bf k}+\varepsilon^o_{-\bf k})a_{-{\bf k}\alpha}a^+_{-{\bf k}\alpha}+
\frac{1}{2}\sum_{{\bf k}\alpha}(\xi_{-\bf k}+\varepsilon^o_{-\bf k}).
\end{eqnarray}
After standard transformation in $H_{int}$ according to mean field approach the Hamiltonian acquires the following form
\begin{eqnarray}
H=\frac{1}{2}\sum_{\bf k}(\xi_{\bf k}+\varepsilon^o_{\bf k})a^+_{{\bf k}\alpha}a_{{\bf k}\alpha}
-\frac{1}{2}\sum_{\bf k}(\xi_{-\bf k}+\varepsilon^o_{-\bf k})a_{-{\bf k}\alpha}a^+_{-{\bf k}\alpha}
\nonumber\\
+\frac{1}{2}\sum_{\bf k}\Delta_{{\bf k},\alpha\beta}a^+_{{\bf k}\alpha}a^+_{-{\bf k}\beta}+
\frac{1}{2}\sum_{\bf k}\Delta^\dagger_{{\bf k},\alpha\beta}a_{-{\bf k}\alpha}a_{{\bf k}\beta}~~~~\nonumber\\
+\frac{1}{2}\sum_{{\bf k}\alpha}(\xi_{-\bf k}+\varepsilon^o_{-\bf k})+\frac{1}{2}\sum_{\bf k}\Delta_{{\bf k},\alpha\beta}F^+_{{\bf k},\beta\alpha},~~~~~~
\label{mf}
\end{eqnarray}
 where the matrix of the order parameter
 \begin{equation}
 \Delta_{{\bf k},\alpha\beta}=-\sum_{{\bf k}^\prime}V_{\beta\alpha,\lambda\mu}({\bf k},{\bf k}^\prime) F_{{\bf k}^\prime,\lambda\mu}
 \label{Delta}
 \end{equation}
 is expressed through "anomalous average"
 $$
 F_{{\bf k},\alpha\beta}=\langle a_{{\bf k}\alpha}a_{-{\bf k}\beta}   \rangle.  
 $$ Here, $\langle ...\rangle$ means subsequent quantum mechanical and thermal averaging.
 
 Introducing operators
 \begin{equation}
 A^+_{{\bf k},i}=(a^+_{{\bf k}\alpha,}a_{-{\bf k}\alpha}),~~~~~A_{{\bf k},i}=\begin{pmatrix} a_{{\bf k}\alpha}\\a^+_{-{\bf k}\alpha}   \end{pmatrix}
 \end{equation}
 one can rewrite Eq.(\ref{mf}) in more compact form
 \begin{eqnarray}
 H=\frac{1}{2}\sum_{\bf k}\varepsilon_{{\bf k},ij}A^+_{{\bf k},i}A_{{\bf k},j}~~~~~\nonumber\\
 +\frac{1}{2}\sum_{{\bf k}\alpha}(\xi_{\bf k}-\varepsilon^o_{\bf k})
 +\frac{1}{2}\sum_{\bf k}\Delta_{{\bf k},\alpha\beta}F^+_{{\bf k},\beta\alpha}.
  \end{eqnarray}
 Here,
 \begin{equation}
 \varepsilon_{{\bf k},ij}=\begin{pmatrix}(\xi_{\bf k}+\varepsilon^o_{\bf k} )\delta_{\alpha\beta} &  \Delta_{{\bf k},\alpha\beta}\\
 \Delta^\dagger_{{\bf k},\alpha\beta}& (- \xi_{\bf k}+\varepsilon^o_{\bf k} ) \delta_{\alpha\beta} \end{pmatrix}.
\end{equation} 
 Diagonalising Hamiltonian by means the Bogolubov transformation
 \begin{equation}
 A_{{\bf k},i}=U_{ij}B_{{\bf k},j},~~~~~~~U_{ij}=\begin{pmatrix}u_{{\bf k},\alpha\beta}&  v_{{\bf k},\alpha\beta}\\
 v^\dagger_{{\bf k},\alpha\beta}& - u_{{\bf k},\alpha\beta}\end{pmatrix},
 \end{equation}
 \begin{eqnarray}
 u_{{\bf k},\alpha\beta}=\frac{\xi_{\bf k}+E_{\bf k}^e}{\sqrt{(\xi_{\bf k}+E_{\bf k}^e})^2+\Delta^2_{\bf k}}\delta_{\alpha\beta},\\
 v_{{\bf k},\alpha\beta}=\frac{\Delta_{\alpha\beta}({\bf k})}{\sqrt{(\xi_{\bf k}+E_{\bf k}^e})^2+\Delta^2_{\bf k}},~~~~
 \end{eqnarray}
 \begin{equation}
 E_{\bf k}^e=\sqrt{\xi_{\bf k}^2+\Delta^2_{\bf k}},~~~~~~~~~\Delta^2_{\bf k}=\frac{1}{2}\mathrm{Tr}\hat\Delta^\dagger({\bf k})\hat\Delta({\bf k}),
 \end{equation}
 we obtain
 \begin{equation}
 \frac{1}{2}\sum_{\bf k}\varepsilon_{{\bf k},ij}A^+_{{\bf k},i}A_{{\bf k},j}=\frac{1}{2}\sum_{\bf k}E_{{\bf k},ij}B^+_{{\bf k},i}B_{{\bf k},j},
 \end{equation}
 where
 \begin{equation}
 E_{{\bf k},ij}=\begin{pmatrix}(\varepsilon_{\bf k}^o+E_{\bf k}^e)\delta_{\alpha\beta} &  0\\
 0& (\varepsilon_{\bf k}^o-E_{\bf k}^e) \delta_{\alpha\beta} \end{pmatrix}.
 \end{equation}
 Thus, the energy of excitations is
 \begin{equation}
 E_{\bf k}=\varepsilon_{\bf k}^o+E_{\bf k}^e.
 \end{equation}
 The corresponding density of states is
 \begin{equation}
 N(E)=2\int\frac{d^3 {\bf k}}{(2\pi)^3}\delta(E-E_{\bf k}).
\end{equation} 
 Any  superconducting  state is {\bf gapless } $N(E=0)\ne 0$. This property of superconducting states in superconductors with toroidal order in particular means nonzero specific heat ratio $(C(T)/T)_{T\to 0}\ne 0$ in completely pure metal without impurities and crystal imperfections.

 The order parameter is determined by Eq.(\ref{Delta})
 \begin{equation}
 \Delta_{{\bf k},\alpha\beta}=-\sum_{{\bf k}^\prime}V_{\beta\alpha,\lambda\mu}({\bf k},{\bf k}^\prime) \langle a_{{\bf k}\lambda}a_{-{\bf k}\mu}   \rangle.
 \end{equation}
 By application to this expression the Bogolubov transformation we obtain
 \begin{eqnarray}
 \Delta_{{\bf k},\alpha\beta}=-\int\frac{d^3 {\bf k}^\prime}{(2\pi)^3}V_{\beta\alpha,\lambda\mu}({\bf k},{\bf k}^\prime)
 \frac{1-f_{{\bf k}^\prime}-f_{{-\bf k}^\prime}}{2E_{{\bf k}^\prime}^e}\Delta_{\lambda\mu}({\bf k}^\prime)\nonumber\\
 =-\int\frac{d^3 {\bf k}^\prime}{(2\pi)^3}V_{\beta\alpha,\lambda\mu}({\bf k},{\bf k}^\prime)
 \frac{\tanh\frac{E_{{\bf k}^\prime}}{2T}+\tanh \frac{E_{{-\bf k}^\prime}}{2T}}
 {4E_{{\bf k}^\prime}^e}\Delta_{\lambda\mu}({\bf k}^\prime).~
 \label{gap}
 \end{eqnarray}
 Here, we used the symmetry properties of the order parameter and expressed the average $\langle b^+_{{\bf k}\alpha}b_{{\bf k}\beta}   \rangle=f_{\bf k}\delta_{\alpha\beta}$ through   the Fermi distribution function
 \begin{equation}
 f_{\bf k}=f(E_{\bf k})=\frac{1}{\exp((\varepsilon_{\bf k}^o+E_{\bf k}^e)/T)+1}.
 \end{equation}

 At $T\to T_c$ one can neglect $\Delta^2_{\bf k}$ in $E_{\bf k}^e$ in Eq.(\ref{gap}). Estimating the integral with logarithmic accuracy we come to
 the expression for critical temperature similar to usual BCS formula
 \begin{equation}
 T_c\approx \varepsilon_0\exp\left (-\frac{1}{\tilde N_0V_{\Gamma}}\right ),
 \end{equation}
where $\varepsilon_0$ is a cut-off  for energy of pairing interaction and $\tilde N_0$ is the density of states averaged over the Fermi surface with a weight corresponding to the angular dependent functions of given irreducible representation.
 
 \section{Conclusion}
In summary, we have developed a theory of superconductivity in metals with toroidal ordering. The treatment is  limited to the case of single band metal with point symmetry groups with a one-dimensional representations. A generalisation for the  multi-band metals and superconducting states corresponding to multi-dimensional representations as well for the metals with more complex multipole order can be performed without much difficulty.
 
 \bigskip
 
 \begin{widetext}
 
\begin{figure}
\includegraphics
[height=.2\textheight]
{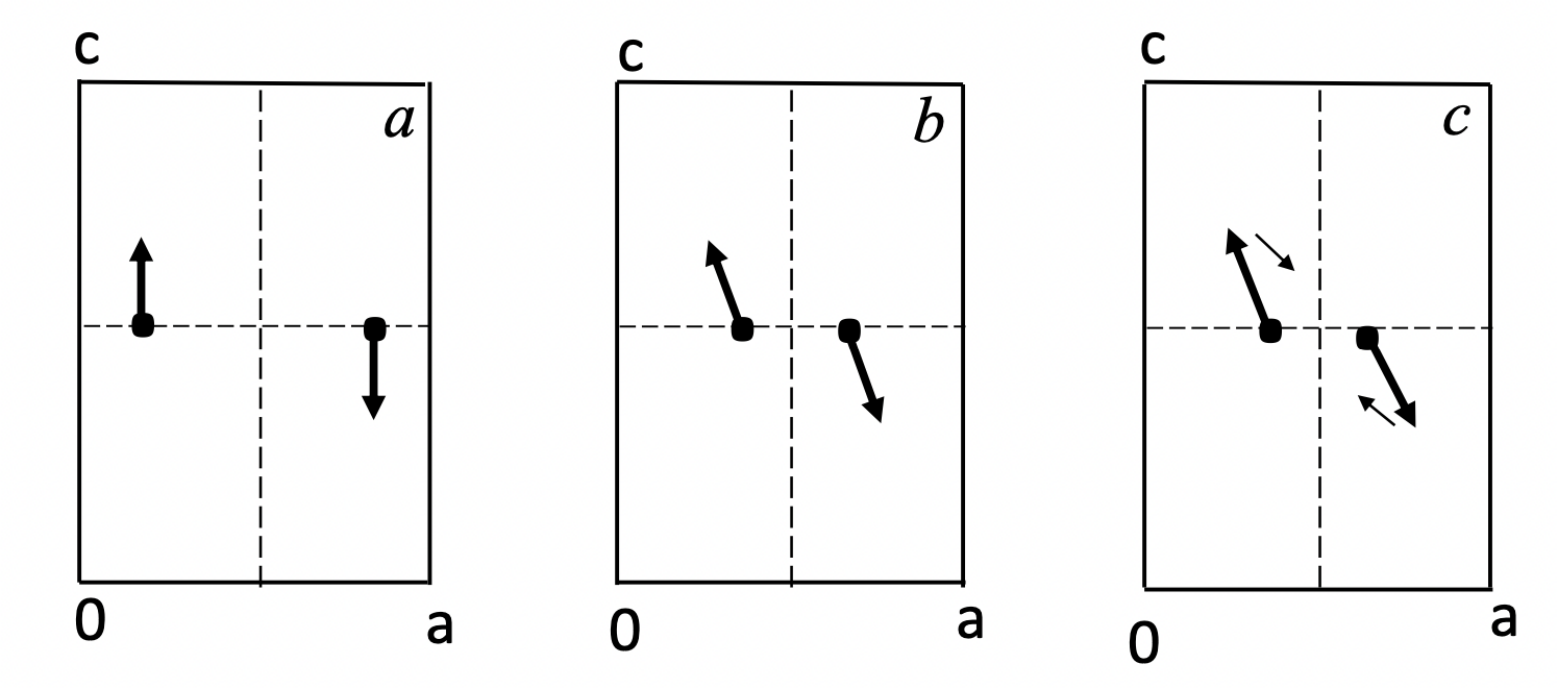}
 \caption{ 
(a) The $y=b/2$ plane  of elementary cell of an antiferromagnet with point  symmetry  group ${\bf D}_{2h}({\bf D}_2)$;
(b) The $y=b/2$ plane  of elementary cell of an antiferromagnet with point symmetry group ${\bf C}_{2h}({\bf C}_2)$;
(c) The $y=b/2$ plane  of elementary cell of an antiferromagnet with point symmetry group ${\bf C}_{2h}({\bf C}_{\sigma_x})$
The moments are tipped out of the plane $y=b/2$ in the opposite directions. 
}
\end{figure}

 \end{widetext}

 \appendix
 
 \section{Superconducting order parameters in metals with toroidal order}

The magnetic structures  corresponding to  point symmetry groups:
${\bf D}_{2h}({\bf C}_{2v})$, ${\bf C}_{2h}({\bf C}_2)$, ${\bf C}_{2h}({\bf C}_{\sigma_x}) $ are sketched on Fig.2 a,b,c.
Below are written the functions of irreducible representations suitable for superconducting states in antiferromagnets with these symmeries.

\subsection{Group ${\bf D}_{2h}({\bf C}_{2v})$}

This group consists of elements 
\begin{equation}
E,C_{2y},\sigma_x,\sigma_z,RC_{2x},RC_{2z},R\sigma_y,RI.
\end{equation}
The odd part of electron band energy is
\begin{equation}
\varepsilon^o_{\bf k}=\gamma\sin k_yb.
\end{equation}
The magneto-electric part of thermodynamic potential has the following form 
$$\Phi_{ME}=-\alpha_{xz}E_xH_z-\alpha_{zx}E_zH_x.$$
The functions of irreducible representations for corresponding superconducting states are
 
\bigskip
 
 \begin{tabular}{|c|c|c|}
\hline
 $\Gamma$& $\phi^s_{\bf k}$&$\mbox{\boldmath$\phi$}^t_{\bf k}$ \\
 \hline
 ~A~& ~$a\hat k_x^2+ b\hat k_y^2+c\hat k_z^2$~&~$ i (\hat k_z\hat x+\hat k_x\hat z)$~\\
 \hline
 ~B$_3$~&~$\hat k_y\hat k_z$~&~$ i(\hat k_y\hat x+\hat k_x\hat y)$~\\
 \hline
  ~B$_2$~&~$\hat k_x\hat k_z$~&~$ i(\hat k_x\hat x+\hat k_y\hat y+\hat k_z\hat z)$~\\
  \hline
 ~B$_1$~&~$\hat k_x\hat k_y$~&~$ i(\hat k_z\hat y+\hat k_y\hat z)$~\\
  \hline
 \end{tabular}
 
 \bigskip 

 One must remember that  each term in each cell may include some amplitude (a real numerical factor or a real function of  squares of  wave vector components). 
 
 \bigskip 

\subsection{Group ${\bf C}_{2h}({\bf C}_2)$}

This group is subgroup of previous one. It consists of elements 
\begin{equation}
E,C_{2y},R\sigma_y,RI.
\end{equation}
The odd part of electron band energy is 
\begin{equation}
\varepsilon^o_{\bf k}=\gamma\sin k_yb+\delta\sin k_xa\sin k_yb\sin k_zc.
\end{equation}
The magneto-electric part of thermodynamic potential has the following form 
$$\Phi_{ME}=-\alpha_{xz}E_xH_z-\alpha_{zx}E_zH_x-\alpha_xE_xH_x-\alpha_yE_yH_y-\alpha_zE_zH_z.$$
The functions of irreducible representations for corresponding superconducting states are

\bigskip
  
 \begin{tabular}{|c|c|c|}
\hline
 $\Gamma$& $\phi^s_{\bf k}$&$\mbox{\boldmath$\phi$}^t_{\bf k}$ \\
 \hline
 ~A~& ~$a\hat k_x^2+ b\hat k_y^2+c\hat k_z^2+d\hat k_x\hat k_z$~&~$ i (\hat k_z\hat x+\hat k_x\hat z+\hat k_x\hat x+\hat k_y\hat y+\hat k_z\hat z)$~\\
 \hline
 ~B~&~$a\hat k_x\hat k_y+b\hat k_y\hat k_z$~&~$ i(\hat k_z\hat y+\hat k_y\hat z+\hat k_y\hat x+\hat k_x\hat y)$~\\
  \hline
 \end{tabular}
 
 \bigskip
 
 Once again: each term in each cell may include some amplitude.

 \subsection{Group ${\bf C}_{2h}({\bf C}_{\sigma_x})$}

This group is another subgroup of the group ${\bf D}_{2h}({\bf C}_{2v})$. It consists of elements 
\begin{equation}
E,\sigma_x,RC_{2x},RI.
\end{equation}
The odd part of electron band energy is 
\begin{equation}
\varepsilon^o_{\bf k}=\gamma\sin k_yc.
\end{equation}
The magneto-electric part of thermodynamic potential has the following form 
$$\Phi_{ME}=-\alpha_{xy}E_xH_y-\alpha_{yx}E_yH_x-\alpha_{xz}E_xH_z-\alpha_{zx}E_zH_x.$$
The functions of irreducible representations for corresponding superconducting states are

\bigskip
  
 \begin{tabular}{|c|c|c|}
\hline
 $\Gamma$& $\phi^s_{\bf k}$&$\mbox{\boldmath$\phi$}^t_{\bf k}$ \\
 \hline
 ~A~& ~$a\hat k_x^2+ b\hat k_y^2+c\hat k_z^2+d\hat k_y\hat k_z$~&~$ i (\hat k_y\hat x+\hat k_x\hat y+\hat k_z\hat x+\hat k_x\hat z)$~\\
 \hline
 ~B~&~$a\hat k_x\hat k_y+b\hat k_x\hat k_z$~&~$ i(\hat k_z\hat y+\hat k_y\hat z+\hat k_x\hat x+\hat k_y\hat y+\hat k_z\hat z)$~\\
  \hline
 \end{tabular}
 
 \bigskip
 Each term in each cell may include some amplitude.

\end{document}